\documentclass[aps,twocolumn,prl,showpacs]{revtex4}
\usepackage{amsfonts}
\usepackage{amsmath}
\usepackage{amssymb}
\usepackage{graphicx}

\begin{document}

\preprint{}
\title{Entanglement and criticality in translational invariant harmonic
lattice systems with finite-range interactions}
\author{R. G. Unanyan and M. Fleischhauer}
\affiliation{Fachbereich Physik, Technische Universit\"{a}t Kaiserslautern, 67653,
Kaiserslautern, Germany}

\begin{abstract}
We discuss the relation between entanglement and criticality in 
translationally invariant harmonic lattice systems with non-randon, finite-range interactions.
We show that the criticality of the system as well 
as validity or break-down of the entanglement area law 
are solely determined by the analytic properties of
the spectral function of the oscillator system, which can easily be computed. 
In particular for finite-range couplings we find  a one-to-one correspondence
between an area-law scaling of the bi-partite entanglement 
and a finite correlation length. This relation is strict in the one-dimensional case and
there is strog evidence for the multi-dimensional case.  
We also discuss generalizations to couplings with infinite range. 
Finally, to illustrate our results, a specific 1D example with nearest and 
next-nearest neighbor coupling
is analyzed.
\end{abstract}

\date{\today}
\startpage{1}
\maketitle

%%%%%%%%%%%%%%%%%%%%%%%%%%%%%%%%%%%%%%%%%%%%%%%%%%%%%%%%%%%%%%%%%%%%%%%%%%%%

%%%%%%%%%%%%%%%%%%%%%%%%%%%%%%%%%%%%%%%%%%%%%%%%%%%%%%%%%%%%%%%%%%%%%%%%%%%%

Due to the development of powerful tools to quantify entanglement
there is a growing interest in the relation between entanglement and
criticality in quantum many-body systems. For a variety of spin
models it was shown that in the absence of criticality, there is a strict
relation between the von-Neumann entropy of a compact sub-set of 
spins in the ground state and the 
surface area of the ensemble. E.g. it was
shown in \cite{GVidal-PRL-2003,Korepin,Calabrese,Mezzadri} that the entanglement in non-critical
one-dimensional spin chains approaches a constant value, while 
it grows logarithmically in the critical case, where the correlation 
length diverges. Employing field theoretical methods it was argued 
that in $d$ dimensions the entropy grows as a 
polynomial of power $d-1$ under non-critical conditions, thus 
establishing an area theorem.
A similar relation was suggested for harmonic lattice models in \cite%
{Bombelli-PRD-1986} and \cite{Srednicki-PRL-1993}. Very recently, employing
methods of quantum information for Gaussian states, Plenio \textit{et al. }%
\cite{Plenio} gave a derivation of the area theorem for harmonic
lattice models with nearest-neighbor couplings. All these findings 
suggest a general correspondence between entanglement and criticality
for non-random potentials.
Yet recently special cases have been found for spin chains with Ising-type
interactions \cite{Duer-PRL-2005} and for harmonic lattices systems
\cite{Eisert-preprint} where the correlation length diverges but the
entanglement obeys an area law. Thus the relation between entanglement scaling
and criticality remains an open question. It should also be noted that in 
disorderd systems, i.e. systems with random couplings the relation between
entanglement area law and criticality is broken. 

In the present paper we show that
for harmonic lattice systems with {\it translational invariant}, {\it non-random}, and {\it finite-range}
couplings both
entanglement scaling and criticality
are determined by the analytic properties of the so-called spectral function. For finite-range interactions 
we find that the properties of the spectral function lead to 
a one-to-one correspondence between 
entanglement and criticality.  To
illustrate our results we discuss a specific one-dimensional example with
nearest and next-nearest couplings. Despite the
finite range of the coupling this model undergoes a transition from area-law
behavior to unbounded logarithmic growth of entanglement.

%%%%%%%%%%%%%%%%%%%%%%%%%%%%%%%%%%%%%%%%%%%%%%%%%%%%%%%%%%%%%%%%%%%
%\paragraph{Spectral function:}
%%%%%%%%%%%%%%%%%%%%%%%%%%%%%%%%%%%%%%%%%%%%%%%%%%%%%%%%%%%%%%%%%%%

Let us first consider a one-dimensional system, i.e. a chain of $N$ harmonic
oscillators described by canonical variables $\left( q_{i},p_{i}\right) $, $%
i=1,2,...N$. 
The oscillators are coupled by a translational invariant quadratic
Hamiltonian
\begin{equation}
H=\frac{1}{2}{\displaystyle\sum\limits_{i=1}^{N}}p_{i}^{2}+\frac{1}{2}{%
\displaystyle\sum\limits_{i,j=1}^{N}}V_{ij}q_{i}q_{j}  \label{Hamiltonian}
\end{equation}%
where $V$ is a real, non-random, symmetric matrix with positive eigenvalues. 
 For a 
translational invariant system $V$ is a Toeplitz
matrix, i.e. its elements depend only on the
difference of the indexes $V_{ij}\equiv V_{k}=V_{-k}$. For a finite system 
translational invariance implies furthermore periodic 
boundary conditions $V_{k}=V_{N-k}$. 
We assume in the following that the interactions are of finite range, i.e
that $V_k\equiv 0$ for $k\ge R$, where $R$ is a finite number independent on $N$.
As we will show at the end of the paper some generalizations to infinite range
couplings are possible.
Being positive definite, $V$ has a unique positive
square root $V^{1/2}$ and its inverse $V^{-1/2}$
which completely determine the
ground-state in position (or momentum) representation,
$\Psi_0(Q)\sim (\det V^{1/2})^{1/4}\exp\{-\frac{1}{4}\langle Q|V^{1/2}|Q\rangle\}$, 
\cite{Bombelli-PRD-1986,Srednicki-PRL-1993}.

The most important characteristics of the oscillator system 
is the spectrum of $V$.
Since $V$ is a ciculant matrix its eigenvalues can be expressed in terms of complex roots 
of unity $z_{j}=\exp\{i2\pi j/N\}=\exp\{i \theta_j\}$, $(j=1,\dots ,N)$:
\begin{eqnarray}
\lambda _{j} &=&\sum_{k=0}^{R-1}\,V_{k}\,\left( z_{j}\right)
^{k}=\frac{1}{2}V_0+\frac{1}{2}\sum_{k=-(R-1)}^{R-1}\!\!\! V_{k}\,\,
\left( \mathrm{e}^{i\frac{2\pi
}{N}j}\right) ^{k}.
\label{lambda_j}
\end{eqnarray}
Eq.(\ref{lambda_j}) together with the positivity of $V$ permits the
representation $\lambda
_{j}=h^{2}(z_{j})=|h^{2}(z_{j})|$ where $h(z_{j})$ is a polynomial of order $%
(R-1)/2$ in $z_j$ (assuming for simplicity that $R$ is an odd number). 
Thus
$|h(z)|\sim\prod_{l=1}^{(R-1)/2}\bigl|z-\tilde{z}%
_{l}\bigr|.$ 
where $\tilde{z}_{l}\equiv \exp
\{i\alpha _{l}\}$ are the zeroth of $h(z)$ which are either real or
complex conjugate pairs with $|\tilde{z}_{l}|\geq 1$ \cite{Polya}. 
Let $Q\le R$ be the number of real zeroth $\tilde{z}_{r}$ with multiplicity $m_{r}\in
\{0,1,\dots\}$. Then 
$\prod_{r=1}^{Q}\bigl|z_{j}-\tilde{z}_{r}\bigr|^{m_{r}}=\prod_{r=1}^{Q}%
\bigl(2-2\cos (\theta _{j}-\alpha _{r})\bigr)^{m_{r}/2},$ and
\begin{equation}
\lambda(z_j)=\lambda_j=
\lambda _{0}(z_{j})\prod_{r=1}^{Q}\Bigl(2-2\cos (\theta
_{j}-\alpha _{r})\Bigr)^{m_{r}}.  \label{eigenvalues-2}
\end{equation}%
$\lambda(z)$ is the so-called spectral function.
$\lambda_{0}(z)$ is called the regular part of $\lambda$. It is a polynomial of the complex 
variable $z$ which has zeroth outside the unit circle. As a consequence
its inverse $\lambda_0^{-1}(z)$ is analytic on and inside the unit circle.
$\prod_{r=1}^{Q}
\bigl(2-2\cos (\theta-\alpha _{r})\bigr)^{m_{r}}$ 
is called the singular part. If $\lambda$ is singular, i.e. if in the thermodynamic limit
$V$ has eigenvalues arbitrarily close to zero, the total Hamiltonian, eq.(\ref{Hamiltonian}), has a vanishing
energy gap between the ground and first excited state.

To evaluate sums of eigenvalues in the limit $%
N\to \infty$, one can interpret eq.(\ref{lambda_j}) as a
Fourier series of $\lambda(\theta)$. 
Thus 
$V_k=\frac{1}{2\pi} \int_0^{2\pi}\!\!\mathrm{d}\theta\, \lambda(\theta)\,
\mathrm{e}^{-i\theta k}$. This integral representation is also valid for 
a finite number $N$ of oscillators up to an error ${\cal O}(1/N)$ 
as long as $k\le (N+1)/2$. 
Due to the periodic boundary conditions  $V^{\pm 1/2}$
are also Toeplitz matrices, and their elements $V^{\pm 
1/2}_{ij} =V^{\pm 1/2}_k$ can be
expressed in terms of $\lambda^{\pm 1/2}$ for $k\le (N+1)/2$
\begin{equation}
\left(V^{\pm 1/2}\right)_k =\frac{1}{2\pi}\int_0^{2\pi}\!\!\mathrm{d}\theta \,
\lambda^{\pm 1/2}(\theta)\, \mathrm{e}^{-i\theta\, k}.
\label{Vpm}
\end{equation}
%
%

%%%%%%%%%%%%%%%%%%%%%%% correlation length %%%%%%%%%%%%%%%%%%%%%%%%%%%%%

Since for the spatial correlation of an oscillator system holds $\langle q_i q_{i+l}\rangle
\sim V_l^{-1/2}$ \cite{Reznik}, the analytic properties of $\lambda^{-1/2}$ determine the
spatial correlation length $\xi$: 
\begin{eqnarray}
\xi ^{-1} &\equiv & -\lim_{l\rightarrow \infty }
\frac{1}{l}\ln \left|\langle q_i q_{i+l}\rangle\right| 
= -\lim_{l\rightarrow \infty }\frac{1}{l}\ln
\left\vert V_{l}^{-1/2}\right\vert \nonumber\\
&=&
-\lim_{l\rightarrow \infty }\frac{1}{l}\ln \left\vert \frac{1}{%
2\pi }\int_{0}^{2\pi }\!\!\mathrm{d}\theta \,\lambda ^{-1/2}(\theta )\,%
\mathrm{e}^{-i\theta l}\right\vert .
\end{eqnarray}
If some derivative of $\lambda^{-1/2}(\theta )$, say the $m$th one, does not exists, partial 
integrations shows that the integral has a
contribution proportional to $l^{-m}$. In this case the correlation length $
\xi $ is infinite, defining a critical system. 
If on the other hand $\lambda^{-1/2}(\theta )$ is smooth 
the integral decays faster than any polynomial in $l^{-1}$. In this case 
the correlation length is finite, corresponding to a non-critical system.
From the form of $\lambda(\theta)$ given in eq.(\ref{eigenvalues-2}) it is
clear that a regular spectral function implies a finite correlation length, i.e. 
a noncritical behavior and a singular one an infinite correlation length, i.e. a 
critical behavior.

%%%%%%%%%%%%%%%%%%%%%%% entanglement %%%%%%%%%%%%%%%%%%%%%%%%%%%%%%%

In the following we will show that the analytic properties
of $\lambda$
also determine the entanglement scaling
of the oscillator system. The bi-partite entanglement of a compact block 
of $N_1$ oscillators (inner partition ${\cal I}$)
with the rest (outer partition ${\cal O}$) is determined 
by the $N_1$-dimensional sub-matrices $A$ and $D$ 
 \cite{Bombelli-PRD-1986,Srednicki-PRL-1993,Reznik,Plenio}
\begin{equation}
V^{-1/2}=\left[
\begin{array}{cc}
A & B \\
B^{T} & C%
\end{array}%
\right] ,\qquad V^{1/2}=\left[
\begin{array}{cc}
D & E \\
E^{T} & F%
\end{array}%
\right] ,  \label{BlockForm}
\end{equation}
$C$ and $F$ are here $\left( N-N_{1}\right)
\times \left( N-N_{1}\right) $ matrices.
The entropy is given by the eigenvalues $\mu
_{i}\geq 1$ of the matrix product $A\cdot D$ \cite{Plenio}:
\begin{eqnarray}
S= \sum_{i=1}^{N_1} f\left(\sqrt{\mu_i}\right),\label{Entropy}
\end{eqnarray}
where $f(x) = \frac{x+1}{2}\ln\frac{x+1}{2} - \frac{x-1}{2}\ln\frac{x-1}{2}$.
Despite the simplicity of its form, (\ref{Entropy}) cannot be
evaluated in general. This is in contrast to spin systems where $A\cdot D$
is itself a Toeplitz matrix \cite{Korepin,Mezzadri}.

%%%%%%%%%%%%%%%%%%%%% entropy upper bound %%%%%%%%%%%%%%%%%%%%%%%%%%%%%%%%%%

An {\it upper bound} to $S$ can be found from the logarithmic negativity
$\ln||\rho^\Gamma||$, where $\rho^\Gamma$ is the partial transpose of the
total ground state $\rho$ and $||\cdot||$ denotes the trace norm.
As shown in \cite{Plenio,Eisert-preprint} the logarithmic negativity is bounded by
the square root of the maximum eigenvalue of $V$ and a sum of absolute values
of matrix elements of $V_{ij}^{-1/2}$ between all sites $i \in {\cal I}$ and $j\in {\cal O}$.
\begin{equation}
S \le 4 \lambda^{1/2}_{\rm max}
\sum_{i\in{\cal I}}\sum_{j\in{\cal O}}\left|V^{-1/2}_{ij}\right|.
\label{negativity}
\end{equation}
%
%
%%%%%%%%%%%%%%%%%%%%  entropy lower bound %%%%%%%%%%%%%%%%%%%%%%%%%%
%
A {\it lower bound} to the entropy can be found making use of
 $\frac{x+1}{2}\ln \frac{x+1}{2}-\frac{x-1}{2}\ln \frac{x-1}{2}%
>\ln x$ . This yields
\begin{equation}
S>\frac{1}{2}{\displaystyle\sum\limits_{i=1}^{N_{1}}}\ln \mu _{i}=\frac{1}{2}%
\ln \Bigl(\det A\cdot D\Bigr).  \label{S-estimate}
\end{equation}
This estimate has a simple and very
intuitive meaning. To see this we first note that the matrix $D$
can be expressed in the form $D=(A-B\cdot C\cdot B^{\top })^{-1}$. Thus
\begin{eqnarray}
S &>&-\frac{1}{2}\ln \det \left( \mathbf{1}-B\cdot C^{-1}\cdot B^{T}\cdot
A^{-1}\right)  \notag \\
&=&-\frac{1}{2}\ln \det \left(
\begin{array}{cc}
A & B \\
B^{T} & C%
\end{array}%
\right) \cdot \left(
\begin{array}{cc}
A^{-1} & 0 \\
0 & C^{-1}%
\end{array}%
\right)  \label{classical} \\
&=&\frac{1}{2}\ln \frac{\det A\,\det C}{\det V^{-1/2}}=\frac{1}{2}\ln \frac{%
\det F\,\det D}{\det V^{1/2}}.  \notag
\end{eqnarray}
where the last equation was obtained by expressing $A$ in terms of $D,E$%
, and $F$. The last line of (\ref{classical}) is just
Shannon's classical mutual information $I(Q_{1}:Q_{2})$ or $I(P_{1}:P_{2})$
respectively, where $Q_{1}=(q_{1},q_{2},\dots ,q_{N_{1}})$ and $%
Q_{2}=(q_{N_{1}+1},\dots ,q_{N})$ are the position vectors of the two
subsystems and $P_{1,2}$ the respective momentum vectors. $I(Q_{1}:Q_{2})$
is defined as
\begin{equation}
I(Q_{1}:Q_{2})=\int \mathrm{d}^{N}Q\,p(Q_{1},Q_{2})\,\ln \frac{p(Q_{1},Q_{2})%
}{p_{1}(Q_{1})p_{2}(Q_{2})}
\end{equation}%
where $p(Q_{1},Q_{2})=|\Psi _{0}|^{2}$ is the total and  $p_{1,2}(Q_{1,2})$
the reduced probability density in position space. 
Straight forward calculation shows
\begin{equation}
I\left( Q_{1}:Q_{2}\right)=\frac{1}{2}\ln \frac{\det A\cdot \det C}{\det
V^{-1/2}} \le S.  \label{MutualGaus}
\end{equation}

%%%%%%%%%%%%%%%%%%%%%%%%%%%%%%%%%%%%% %%%%%%%%%%%%%%%%%%

In order to evaluate Shannon's mutual information 
in the form given in eq.(\ref{S-estimate})
we want to make use of the asymptotic properties of Toeplitz
matrices. For this we note that since $V^{\pm 1/2}$ are Toeplitz
matrices, so are $A$ and $D$.
Their elements $A_{k}$, and $D_{k}$ can be obtained from $\lambda ^{\pm 1/2}$
by (\ref{Vpm}) if  $N_1\le (N+1)/2$.

%%%%%%%%%%%%%%%%%%%%%%%%% regular lambda %%%%%%%%%%%%%%%%%%%%%%%%%%%%%

If $\lambda(\theta )$ is {\it regular}, we can apply the strong Szeg\"{o}
theorem \cite{Szegoe}, which states:
\begin{eqnarray}
\det (D)\,  \rightarrow  \,\exp \Bigl\{c_{0}N_1+\sum_{k=0}^{\infty
} k |c_{k}|^{2}\Bigr\},
\end{eqnarray}
for $N_1\to \infty$.
Here the $c_{k}$ are Fourier-coefficients of $\ln \lambda^{-1/2}(\theta)$, 
i.e. 
$c_{k}=\frac{1}{2\pi }\int_{0}^{2\pi }\!\!\mathrm{d}\theta \,\,\ln \lambda
^{1/2}(\theta )\,\mathrm{e}^{-i\theta \,k}$.
Noting that the corresponding coefficients for $A$
have opposite sign, we find the lower bound
\begin{equation}
S\ge \frac{1}{2}\ln(\det(A))+\frac{1}{2}\ln(\det(D))
=\sum_{k=0}^{\infty }\,k\,|c_{k}|^{2}.\label{lower-bound-1-d}
\end{equation}
To find an upper bound to $S$ we make use of eq.(\ref{negativity}).
For a finite-range interaction  there is always a maximum eigenvalue $\lambda_{\rm max}^{1/2}$. 
Furthermore since  $\lambda^{-1/2}_0(\theta)$ is smooth,  eq.(\ref{Vpm})
implies an exponential bound to the matrix elements of $V^{-1/2}$.
I.e. $|V^{-1/2}_{ij}| \le K \exp\{-\alpha|i-j|\}$, for $|i-j|\le (N+1)/2$, where
$K,\alpha >0$. With this
we find
\begin{eqnarray}
\sum_{i\in{\cal I}}\sum_{j\in{\cal O}}
\left|V^{-1/2}_{ij}\right| &=& 2 N_1 \sum_{k=N_1+1}^{(N+1)/2} |V_k^{-1/2}|
+ 2 \sum_{k=1}^{N_1} k |V_k^{-1/2}|\nonumber\\
&<& \frac{2 K {\rm e}^{-\alpha}}{(1-{\rm e}^{-\alpha})^2}
\end{eqnarray}
for $N,N_1\to\infty$. Thus $S$ has also a finite upper bound in 1D.
One recognizes 
that for one dimensional harmonic chains with a {\it regular} spectral
function $\lambda(\theta)$ the entropy has a lower and an upper 
bound independent on the number of
oscillators, which implies an area theorem. 
Furthermore, as shown above, the spatial correlation length is
finite, i.e. the system is non-critical.

Let us now consider a {\it singular} function $\lambda$.
In this case we can calculate the asymptotic behavior of the 
Toeplitz determinants
using Widom's theorem \cite{Widom}. This theorem states that 
for $N_1\to\infty $ and for $m_r>-1$:
\begin{equation}
\det D \, \rightarrow\, \exp\left\{c_0 N_1\right\} \, N_1^{\sum_r m_r^2/4}.
\end{equation}
 Widom's theorem cannot be applied to $A$, since $\lambda^{-1/2}(\theta)\sim \prod_r
(2-2\cos(\theta-\alpha_r))^{-m_r/2}$ involves negative exponents. We 
thus employ
the alternative expression (\ref{classical}) containing the matrices $D$ and
$F$. Since the elements of $D$ and $F$ 
can only be obtained by the Fouriertransform (\ref{Vpm})
if their dimension is at most $(N+1)/2$, 
there is only one particular decomposition which we can consider, namely
$N_1=(N-1)/2$ and $N_2=(N+1)/2$. For the same reason it is not
possible to apply Widom's theorem to $V^{1/2}$ as a whole. $\det V^{1/2}$ can however easily 
be calculated directly from the discrete eigenvalues (\ref{eigenvalues-2}).
After a lengthy but straight forward calculation we eventually obtain the 
following expression for the mutual
information with $N_1=(N-1)/2$ and $N_2=(N+1)/2$ 
\begin{eqnarray}
I &=& 
\left(\sum_{r=1}^Q \frac{m_r^2}{4}\right) \ln N +
\mathrm{const.}\, \, . \,
\end{eqnarray}
Thus a singular spectral function $\lambda^{1/2}(\theta)$ in the case of half/half partitioning 
leads to a lower bound to the entropy that grows logarithmically with
the number of oscillators stating a break-down of the area law of
entanglement. As shown above a singular spectral function also implies a diverging
spatial correlation length, defining a critical system.

\

The above discussion can be extended to $d$ dimensions. In this case one would consider the
entropy $S$ of a hypercube of oscillators with dimensions $N_{1}\times
N_{2}\times \cdots \times N_{d}$. Since we are interested in the thermodynamic
limit we can again assume $N_i\le (N+1)/2$.
In this case the matrices $A$ and $D$ are Toeplitz
matrices with respect to each spatial direction and their elements 
$A_{k_{1},k_{2},\dots ,k_{d}}$ can be obtained from the square root of the 
$d$-dimensional function %
$\lambda (\theta _{1},\dots ,\theta
_{d})=\sum_{k_{1}=0}^{N_{1}-1}\dots
\sum_{k_{d}=0}^{N_{d}-1}V_{k_{1},\dots ,k_{d}}\,\exp \left\{
i\sum_{j=1}^{d}\theta _{j}\,k_{j}\right\}$. If $\lambda ^{1/2}$ is
{\it regular}, the $d$-dimensional Szeg\"{o} theorem holds \cite{d-dimSzegoe}, which asserts that the
Toeplitz determinant of dimension $n_{1}\times n_{2}\times \cdots \times
n_{d}$ has the asymptotic form
\begin{eqnarray}
\det D \rightarrow \exp \Bigl\{c_{0}n_{1}\cdots n_{d}+\sum_{j=1}^{d}\frac{%
n_{1}\cdots n_{d}}{n_{j}}|C_{j}|\Bigr\},
\end{eqnarray}
where $c_{0}=\frac{1}{(2\pi )^{d}}\int_{0}^{2\pi }\!\!\mathrm{d}\theta _{1}\cdots
\int_{0}^{2\pi }\!\!\mathrm{d}\theta _{d}\,\ln \Bigl(\lambda ^{1/2}(\theta
_{1},\dots ,\theta _{d})\Bigr)$,
and the $C_{i}$ are some constants, whose explicit form is of no interest
here. We see that under the above conditions for the $d$-dimensional
characteristic function $\lambda ^{1/2}(\theta _{1},\dots ,\theta _{d})$ the
entropy has the lower bound
\begin{equation}
S>\sum_{j=1}^{d}\,\frac{n_{1}n_{2}\cdots n_{d}}{n_{j}}\,\,C_{j}\,\sim
\,n^{d-1}\label{lower-bound-d}
\end{equation}%
which is again proportional to the surface area.
We note that the lower bound (\ref{lower-bound-d}) to the entropy given by the
multi-dimensional Szeg\"o theorem is more general than the estimates
given in \cite{Plenio} and 
\cite{Eisert-preprint}, which are restricted to nearest neigbor
interactions. 
From the exponential bound to the matrix elements
of $V^{-1/2}$ one can also
find an upper bound to the entropy using eq.(\ref{negativity})
\begin{eqnarray}
\sum_{i\in{\cal I}}\sum_{j\in{\cal O}}
\left|V^{-1/2}_{ij}\right| 
\le  K \sum_{j=1}^{d}\,\frac{n_{1}n_{2}\cdots n_{d}}{n_{j}}\,\sim
\,n^{d-1}
.\label{upper-bound-d}
\end{eqnarray}
Eqs.(\ref{lower-bound-d}) and (\ref{upper-bound-d}) establish an area law
for arbitrary dimensions in the case of a regular spectral function.

%%%%%%%%%%%%%%%%%%%%%%%%%%%   singular %%%%%%%%%%%%%%%%%%%%%%%%%%%%%%%%%%%%

In order to obtain a lower bound to the entropy for a {\it singular} spectral
function in more than one dimension and to show a corresponding break-down of the
entanglement area law, one would need a multi-dimensional generalization 
of Widoms theorem \cite{Widom}. Although no such generalization is known to us,
there is strong evidence for a break down of the area law in higher dimensions. 
First of all for an interaction matrix that
is separabel in the $d$ dimensions, i.e. whose elements can be written as
products $V_{i_1,j_1}\, V_{i_2,j_2}\, \dots \, V_{i_d,j_d}$, the 1D
discussion can straight-forwardly extended to $d$ dimensions. Secondly Widom has
given a generalization of his matrix theorem to operator functions $f(A)$
on $R^d$ \cite{Widom-2}. The proof given in \cite{Widom-2} makes however use of  
strong conditions on $f$ that are not fulfilled for the case we are interested
here.

%%%%%%%%%%%%%%%%%%%%%%%%%%%%%%%%%%%%%%%%%%%%%%%%%%%%%%%%%%%%%%%%%%%%%%%%
%\paragraph{An example:}
%%%%%%%%%%%%%%%%%%%%%%%%%%%%%%%%%%%%%%%%%%%%%%%%%%%%%%%%%%%%%%%%%%%%%%%%

\begin{figure}[htb]
\begin{center}
\includegraphics[width=7.0cm]{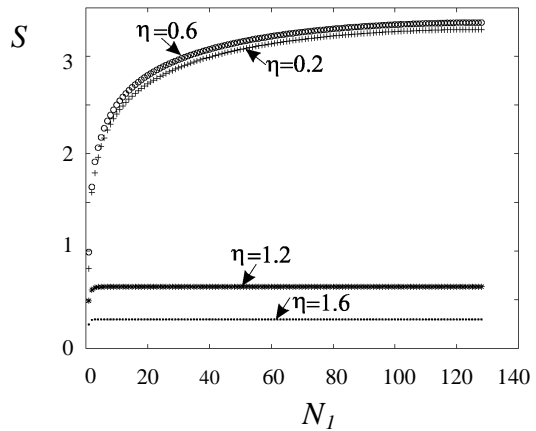}
\caption{Entropy as function of partition size for
non-critical ($\eta=1.2, 1.6$) and critical harmonic chain ($\eta=0.2, 0.6$) 
for the example of the text obtained from numerical calculation of
eq.(\ref{Entropy}).}
\label{mutual-information}
\end{center}
\end{figure}

To illustrate validity and break-down of the area theorem let us 
consider the Hamiltonian
$H=\frac{1}{2}\sum_{i=1}^{N}p_{i}^{2} + \frac{1}{2}\sum_{i=1}^{N}\left( -2\eta
q_{i}+q_{i+1}+q_{i-1}\right) ^{2}  
$
with periodic boundary conditions. 
The square
root of the spectral function reads in this case $\lambda
^{1/2}(\theta )=\left\vert 2\eta -2\cos \theta \right\vert $. For $\eta >1$,
$\lambda^{1/2}$ is regular and the
correlation length is finite. For $\eta <1$, $\lambda ^{1/2}(\theta )$
can be written as $\lambda ^{1/2}(\theta )=\left( 2-2\cos \left( \theta
+\theta _{0}\right) \right) ^{1/2}\left( 2-2\cos \left( \theta -\theta
_{0}\right) \right) ^{1/2}$, with $\eta =\cos \theta _{0}$, and thus is
singular. In this case the correlation length is infinite. 
We have numerically calculated the entropy for this system 
for different values of $\eta$. The results are shown in 
fig.\ref{mutual-information}. One recognizes an
unlimited logarithmic growth of $S$ for $\eta <1$ and a saturation
for $\eta>1$.

%%%%%%%%%%%%%%%%%%%%%%%%%%%%%%%%%%%%%%%%%%%%%%%%%%%%%%%%%%%%%%%%%%%%%%%%%

%%%%%%%%%%%%%%%%%%%%%%%%%%%%%%%%%%%%%%%%%%%%%%%%%%%%%%%%%%%%%%%%%%%%%%%%%%

In the present paper we discussed the relation between entanglement and
criticality in translational invariant harmonic lattice
systems with finite-range couplings. We have shown that upper and lower bounds to the entropy
of entanglement as well as the correlation length are solely determined
by the analytic properties of the spectral function. If the spectral function
is regular, the entanglement obeys an area law and the system is
non-critical. If the spectral function has a singular part, the
area law breaks down and the system is critical. Thus
for harmonic lattice systems with {\it translational invariant}, {\it non-random} and
{\it finite-range} couplings there is a one-to-one correspondence between 
entanglement and criticality. We note that some of our results apply also to
more general couplings. For the estimates of the entropy it is
sufficient that the number of roots of $h(z)$ on the unit circle is finite.
This is always fulfilled for banded coupling matrices $V$ but also holds under more
general conditions. For couplings of infinite range, the regular part of the
spectral function $\lambda_0$ is no longer a polynomial. Thus $\lambda_0^{-1/2}$ may not be smooth anymore
and could have a singularity in a derivative of some order. In such a case the spectral function
could be regular, allowing for an entanglement area theorem, and at the same time the correlation
length would be infinite, i.e. the system would be critical as in the example of Ref.\cite{Eisert-preprint}.

The authors would like to thank J. Eisert and M. Cramer for many stimulating discussions.
This work was supported by the DFG through the SPP \textquotedblleft Quantum
Information\textquotedblright\ as well as the European network QUACS.

%%%%%%%%%%%%%%%%%%%%%%%%%%%%%%%%%%%%%%%%%%%%%%%%%%%%%%%%%%%%%%%%%%%%%%%%%%%%%%%%%%


\begin{thebibliography}{99}
\bibitem{GVidal-PRL-2003} G. Vidal, J. I. Lattore, E. Rico, and A. Kitaev,
Phys. Rev. Lett. \textbf{90}, 227902 (2003).

\bibitem{Korepin}  A.R. Its, B.Q. Jin, and V.E. Korepin, J. Math. Phys. A \textbf{38}, 2975 (2005); 
 B.Q. Jin and V.E. Korepin, J. Stat. Physics \textbf{116}, 79 (2004).

\bibitem{Calabrese} P. Calabrese and J. Cardy, J. Stat. Mech. Theory E, P06002 (2004).

\bibitem{Mezzadri} J.P. Keating and F. Mezzadri, Comm. Math. Phys. \textbf{252}, 543 (2004);
Phys. Rev. Lett. \textbf{94}, 050501 (2005).


\bibitem{Bombelli-PRD-1986} L. Bombelli, R. K. Koul, J. Lee, and R. D.
Sorkin, Phys. Rev. D \textbf{34}, 373 \ (1986).

\bibitem{Srednicki-PRL-1993} M. Srednicki, Phys. Rev. Lett. \textbf{71}, 666
(1993).

\bibitem{Plenio} M. B. Plenio, J. Eisert, J. Drei\ss ig, and M. Cramer,
Phys. Rev. Lett. \textbf{94}, 060503 (2005).



\bibitem{Duer-PRL-2005} W. D\"ur, L. Hartmann, M. Hein, M. Lewenstein, and H.-J. Briegel,
 Phys. Rev. Lett. {\bf 94}, 097203 (2005).

\bibitem{Eisert-preprint} M. Cramer, J. Eisert, M.B. Plenio, and J. Drei\ss ig,
preprint quant-ph/0505092.



%\bibitem{Horn} R.A. Horn, C.R. Johnson, \textit{Matrix Analysis}, Cambridge
% University Press, Cambridge, MA, 1985.


% \bibitem{Davis} P.J. Davis, \textit{Circulant Matrices}, 2nd ed. New York:
% Chelsea, 1994.



\bibitem{Polya} G. Polya und G. Szeg\"{o}, \textit{Aufgabe und Lehrs\"{a}tze aus
der Analysis II,} Spinger-Verlag, 1964.


\bibitem{Reznik} A. Botero and B. Reznik, Phys. Rev. A.\textbf{70}, 052329
(2004).




% \bibitem{Shannon} C. Shannon and W. Weaver, \textit{The mathematical theory
% of communication}, University of Illions Press,1972.

\bibitem{Szegoe} U. Grenander and G. Szeg\"{o}, \textit{Toeplitz forms and
their applications}, University of California Press, Berkly 1958.

\bibitem{d-dimSzegoe} I.J. Linnik, Math. USSR Izvestija \textbf{9}, 1323 (1975).
[Izv. Akad. Nauk SSSR, Ser. Mat. Tom 39 (1975)]

\bibitem{Widom} H. Widom, Amer. J. Math. \textbf{95,} 333 (1973).

\bibitem{Widom-2} H. Widom, J. Funct. Anal. \textbf{88}, 166 (1990).


\end{thebibliography}
\end{document}